\documentclass[aps,prl,groupedaddress,showpacs]{revtex4-1}

\usepackage{amssymb}
\usepackage{color,graphicx}

\begin{document}

\title{Solitonic Vortices in Bose-Einstein Condensates}

\author{Marek Tylutki$^{1}$}
\author{Simone Donadello$^{1}$}
\author{Simone Serafini$^{1}$}
\author{Lev P. Pitaevskii$^{1,2}$}
\author{Franco Dalfovo$^{1}$}
\author{Giacomo Lamporesi$^{1}$}
\author{Gabriele Ferrari$^{1}$}

\affiliation{\textit{1 INO-CNR BEC Center and Dipartimento di Fisica, Universit\`a di Trento, 38123 Povo, Italy\\
2 Kapitza Institute for Physical Problems RAS, Kosygina 2, 119334 Moscow, Russia}}

\begin{abstract}
We analyse, theoretically and experimentally, the nature of solitonic vortices (SV) in an elongated Bose-Einstein condensate. In the experiment, such defects are created via the Kibble-Zurek mechanism, when the temperature of a gas of sodium atoms is quenched across the BEC transition, and are imaged after a free expansion of the condensate. By using the Gross-Pitaevskii equation, we calculate the in-trap density and phase distributions characterizing a SV in the crossover from an elongate quasi-1D to a bulk 3D regime. The simulations show that the free expansion strongly amplifies the key features of a SV and produces a remarkable twist of the solitonic
plane due to the quantized vorticity associated with the defect. Good agreement is found between simulations and experiments.
\end{abstract}

\maketitle

\section{Introduction}

Defects are ubiquitous in systems with symmetry breaking phase transitions. Independent domains form wherever the system can choose between two or more degenerate ground states, and defects appear at the borders between such domains. This is also the case of the order parameter of Bose-Einstein condensates (BEC), whose phase can assume arbitrary values. Phase defects of a BEC are, for instance, solitons and vortex lines, which can also be created by phase imprinting \cite{Dobrek99,Leanhardt02} or density engineering \cite{Dutton01}. In a temperature quench across the BEC phase transition they appear as a result of the Kibble-Zurek mechanism (KZM). In fact, if the quench is fast enough, the growing condensate has no time to share a unique phase everywhere and defects are spontaneously generated. KZM provides a relation between the density of defects and the quench rate. It was initially developed for cosmological models, but later generalized for condensed matter systems~\cite{KibbleZurek}. In a recent experiment in Trento~\cite{LamporesiKZM13}, such defect production was observed in an ultra-cold gas of sodium atoms, and the KZ power-law scaling was also measured. The nature of the observed defects was analyzed in \cite{Donadello14}. Similar defects were also observed after phase imprinting in a fermionic Lithium gas~\cite{Yefsah13,Ku14}. Related theoretical studies can be found in Refs.~\cite{DefectDynamics}.

The type of defects in a BEC strongly depends on the geometry of the trapping potential and, more precisely, on its dimensionality. In an axially symmetric elongated harmonic trap, the relevant dimensionless parameter is the ratio between the chemical potential and the transverse harmonic oscillator energy, $\gamma= \mu / \hbar \omega_\perp$, which also fixes the ratio between the transverse size of the condensate and the healing length, $\gamma= R_\perp/2\xi$. If $\gamma\sim1$, the system is in a quasi-1D regime, where the natural stable defects are solitons \cite{Brand02}. For $\gamma\gg1$ solitons are subject to the so-called snake instability and decay into more complex structures with smaller energy \cite{Mateo14}.

In an elongated BEC, the most robust structure is represented by the \emph{solitonic vortex} (SV), which displays features of both a vortex and a soliton as a result of the asymmetric confinement in the directions perpendicular to the vortex line \cite{Brand02,Brand01,Komineas03,ParkerPhDthesis} (see Fig. \ref{fig1:sim_trap}a). The structure of a SV can be qualitatively understood by considering the difference between a vortex line aligned along the axis of axisymmetric BEC and a vortex line in the perpendicular direction. In the former case the density and the velocity fields are both isotropic and the phase of the order parameter varies linearly as a function of the angle around the axis.
In the latter case, instead, density and phase are no longer isotropic around the vortex and the equi-phase surfaces bend and squeeze in a region close to the plane containing the vortex line (see Fig. \ref{fig1:sim_trap}a). In this region the phase gradient, and hence the superfluid velocity, is stronger and the system reduces its energy by depleting the density.  Seen from the regions away from the defect plane, the SV behaves as a solitonic excitation of the BEC, with a local density structure and an overall phase jump. In an elongated BEC, a large phase gradient and a significant density depletion occur when $\gamma$ is small, while in the limit $\gamma\gg1$ a SV reduces to the usual quantized vortex in a bulk condensate (Fig. \ref{fig1:sim_trap}b). In this work we aim to better characterize the properties of a SV and show how a SV can be experimentally observed.

\section{The system}

Our experimental system is based on sodium atoms trapped in a cigar-shaped harmonic trap with frequencies $\omega_{x}/2\pi=13\,\mathrm{Hz}$ and $\omega_{\perp}/2\pi=131\,\mathrm{Hz}$ ($\omega_y=\omega_z=\omega_{\perp}$). RF-forced evaporation is performed to cool the atomic sample across the critical temperature with controlled cooling rates. Depending on the cooling rate a certain number of defects is created in the BEC, in agreement with the KZM, as was reported in~\cite{LamporesiKZM13}. Since our aim here is to study the nature of a single defect, we  choose a quenching ramp of $320$~kHz/s, which produces about one defect on average. With 20-25 million atoms the transition temperature is $T_c\simeq700$ nK (for further experimental details see~\cite{Lamporesi13,Donadello14}). In the experiment we attain the regime where $\gamma = 27$. Once a condensate is produced and cooled well below $T_c$, we release the trapping confinement and allow the gas to expand for a long time-of-flight (120 ms) in a levitating magnetic field gradient. 


\section{Numerical simulation}

We are interested in simulating the dynamics of the expansion of a condensate containing a SV. For this we use the Gross-Pitaevskii equation (GPE),
\begin{equation}
i \hbar\ \frac{\partial}{\partial t} \psi = -\frac{\hbar^2}{2m}\ \nabla^2 \psi + \frac{m}{2}\, (\omega_x^2 x^2 + \omega_\perp^2 r_\perp^2)\, \psi + g |\psi|^2 \psi ~,
\end{equation}
for the condensate wave function at zero temperature, $\psi({\bf x})$, normalized to the particle number: $\int d^3 {\bf x}\, \psi^*({\bf x}) \psi({\bf x}) = N$.  The confining potential is weaker along $x$. The coupling constant $g$ is related to the s-wave scattering length $a_s$ by $g = 4 \pi \hbar^2 a_s / m$.

\begin{figure}
\centering
\includegraphics[width = 0.8\textwidth]{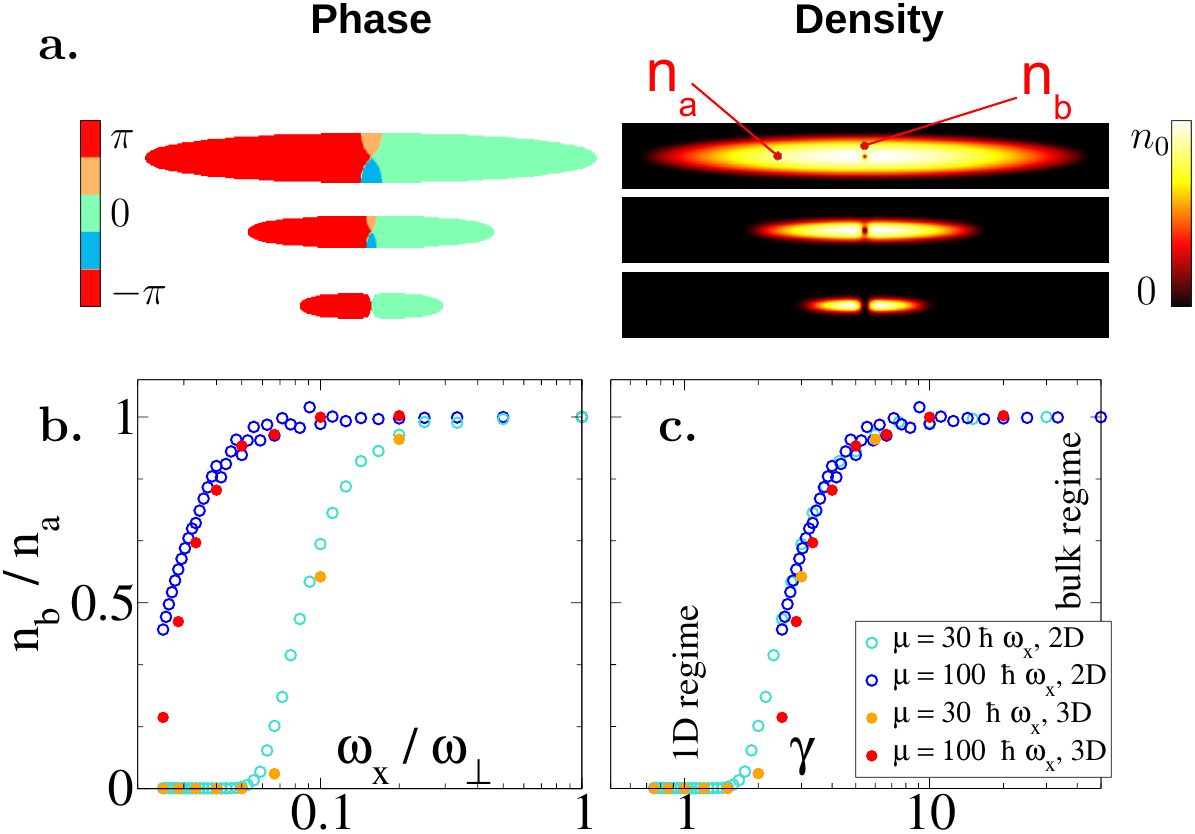}
\caption{ {\bf (a)} Two-dimensional GP simulation for phase and density of elongated BECs with $\gamma=\mu/\hbar\omega_\perp=10,3,1$ (from top to bottom). 
{\bf (b)} Ratio between the density in the solitonic plane ($n_b = n(0,R_\perp / 2)$) and in the bulk BEC ($n_a = n(R_x / 2, 0)$) for two values of the chemical potential $\mu$ as a function of the aspect ratio, both in 2D and 3D BECs. 
{\bf (c)} The same data collapse onto a universal curve when plotted as a function of $\gamma$.}
\label{fig1:sim_trap}
\end{figure}

We start with 2D simulations, which are easier to perform and give some insight into the behaviour of the SV. The condensate is in the $x$-$y$ plane with $\omega_\perp=\omega_y$ and the long axis along $x$. The stationary configuration of the BEC containing a defect is obtained via an imaginary time evolution of the 2D GPE with a suitable phase profile, which is the isotropic phase distribution of a single quantized vortex at $(x=0,y=0)$. Fig. \ref{fig1:sim_trap}a shows the in-trap phase and density profiles after convergence for three values of $\gamma$. Small density depletion in the direction perpendicular to the SV for $\gamma=10$ becomes large for $\gamma=3$ and leads to a dark soliton for $\gamma=1$. The inset of Fig. \ref{fig1:sim_trap}b shows the ratio between densities in different points, $n_b$(0,$\frac{R_\perp}{2}$) and $n_a$($\frac{R_x}{2}$,0), for a BEC with a SV as a function of the aspect ratio $\omega_x/\omega_\perp$ and for three values of chemical potential $\mu$. Here $R_\perp$ and $R_x$ are the Thomas-Fermi radii in the transverse and axial directions respectively. In a spherical trap ($\omega_x/\omega_\perp=1$), the SV becomes a standard vortex whose density depletion is confined in the vortex core of size $\xi$, and for symmetry $n_a = n_b$. For an elongated BEC ($\omega_x/\omega_\perp<1$) the defect crosses over to the SV regime where $0 < n_b / n_a < 1$. Finally, for very elongated traps, the BEC enters the 1D regime, and the defect becomes a dark soliton ($n_b = 0$). The same figure shows that, by plotting the results as a function of $\gamma$, all points collapse onto a single curve, thus confirming that the nature of a defect depends on $\gamma$, in agreement with~\cite{Brand02,Komineas03}.
This analysis quantitatively shows the existence of three regimes, a \emph{1D regime} admitting solitons, a \emph{bulk regime}, with standard vortices, and a \emph{crossover regime} that hosts SVs. Similar results are obtained with full 3D simulations, and the 3D data is only slightly shifted with respect to the 2D points. 

We compare the expansion of 2D condensates containing either a soliton or a SV. In case of a dark soliton the starting phase profile is a uniform phase with a jump $\Delta \varphi = \pi$ across the solitonic line at $(0,y)$. After convergence, the soliton exhibits a density depletion of width $\sim \xi$ at the nodal line.

\begin{figure}
\centering
\includegraphics[width = 0.9\textwidth]{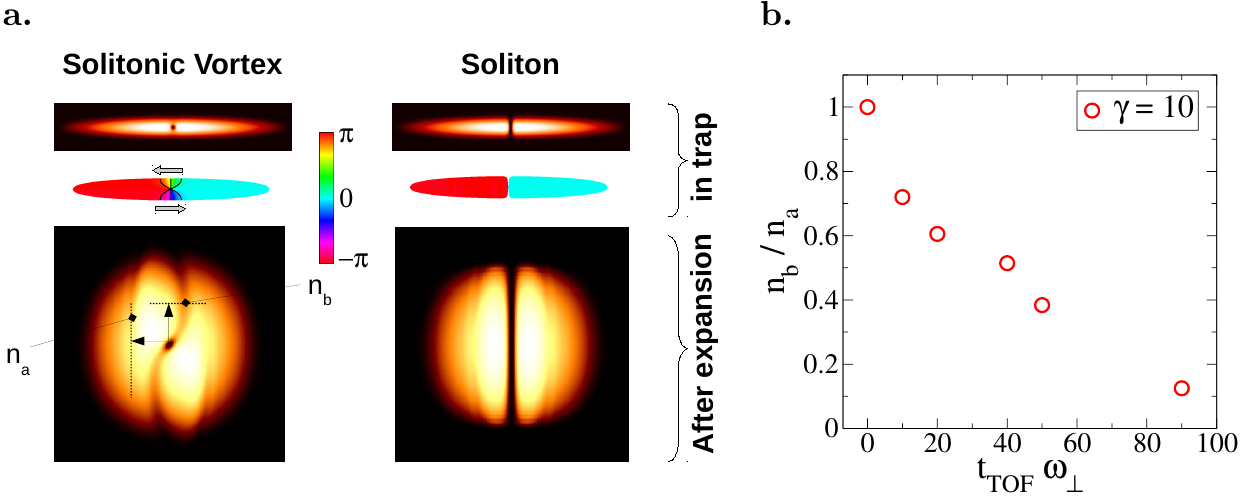}
\caption{ {\bf (a)} 2D simulations of in-trap density (top raw) and phase (middle raw) profiles for an elongated BEC containing a SV (left) or a soliton (right). The density profile calculated after free expansion on one axial oscillation period is shown in the bottom raw. Density color scale is normalized to the peak density of each image.
{\bf (b)} Evolution in time of the density ratio $n_b/n_a$ as in Fig. \ref{fig1:sim_trap} in free expansion for a 3D BEC containing a SV. }
\label{fig2:sim_exp}
\end{figure}

Free expansion is simulated by using the trapped stationary solution as initial state in the real time integration of the GPE, in which $\omega_x$ and $\omega_\perp$ are set to zero. During the expansion, density and mean-field interaction decrease and the healing length (and thus the size of the density depletion of any defect) grows in time. For a soliton this implies that the expanded soliton is wider and can be observed after a long enough time-of-flight. The SV has a more complex phase pattern which, in addition to the widening of the defect, produces two peculiar effects: i) the density depletion in the transverse direction deepens and  ii) the line, along which the density is depleted, twists around the central vortex. This twist comes from the superposition of the velocity fields of the vortex and the velocity of the expansion (or equivalently, the interference of the corresponding phase patterns). This effect lasts as long as the mean-field interaction plays a role. When the expansion becomes ballistic, the twist stops. The combination of i) and ii) makes the SV much more visible after the expansion than in trap. Typical results of simulations in 2D are shown in Fig. \ref{fig2:sim_exp}a. During expansion the visibility of the density depleted plane of the SV is strongly enhanced. 

We perform 3D simulations for comparison with experimental images. In order to reduce the computational time, we limit simulations to $\gamma = 10$ and we dynamically rescale all lengths in the transverse direction according to the Thomas-Fermi scaling law for the expanding order parameter~\cite{Castin96,Massignan03}. Fig. \ref{fig2:sim_exp}b shows $n_b/n_a$ as a function of time. In this case $n_b$ is the density at $(x_m,R_\perp/2)$ with $x_m$ being the position of the local density minimum. The result is consistent with experimental data for which a large $\gamma=27$ would correspond to a negligible depletion in trap but clearly visible after expansion (Fig. \ref{fig3:exp_density}). A detailed comparison of simulation in 2D and 3D shows that the twist is less pronounced in 3D, which is consistent with the fact that in 3D the condensate density decreases faster than in 2D and than the expansion becomes ballistic earlier.

\section{Experimental observations}

\begin{figure}
\centering
\includegraphics[width = .24\textwidth]{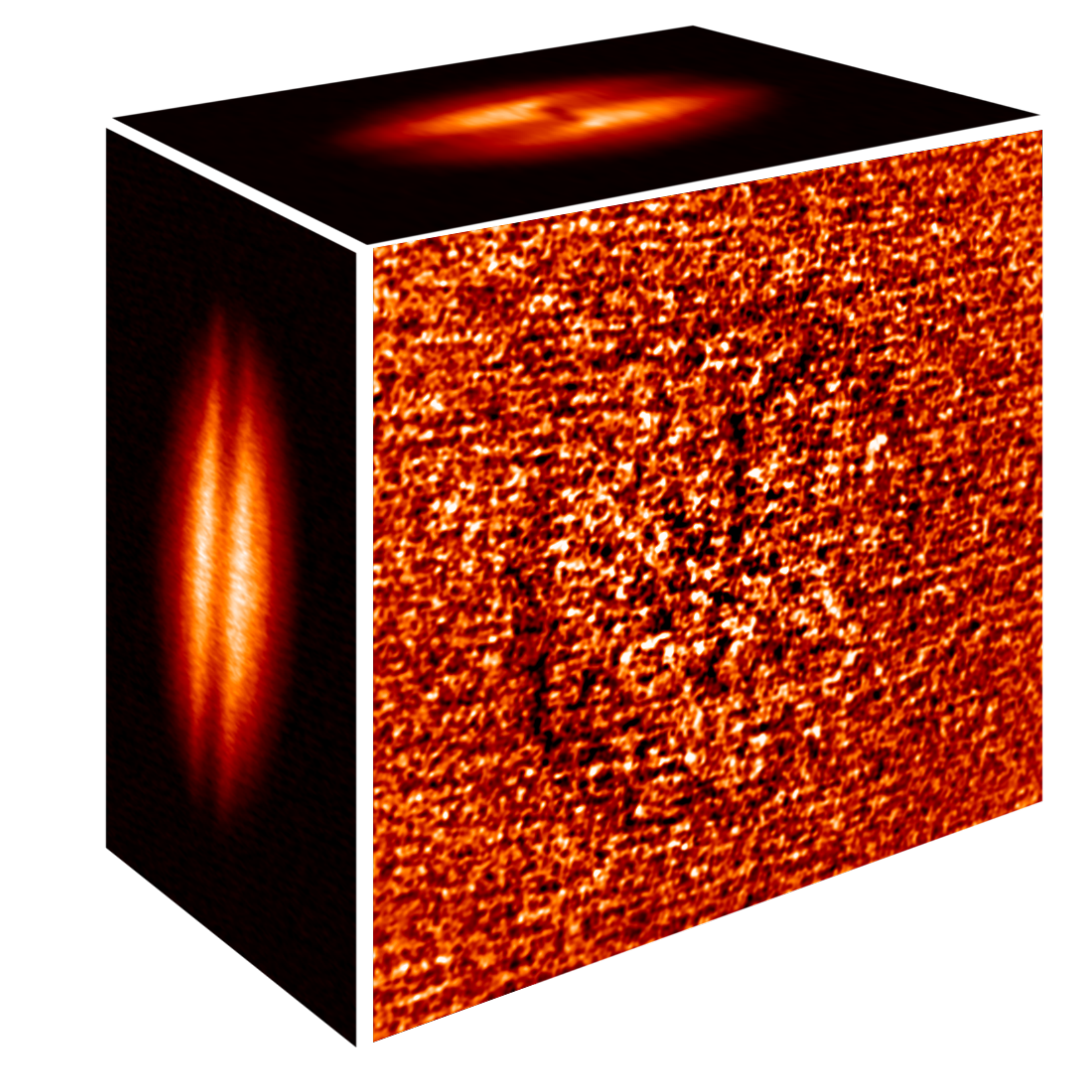}
\includegraphics[width = .24\textwidth]{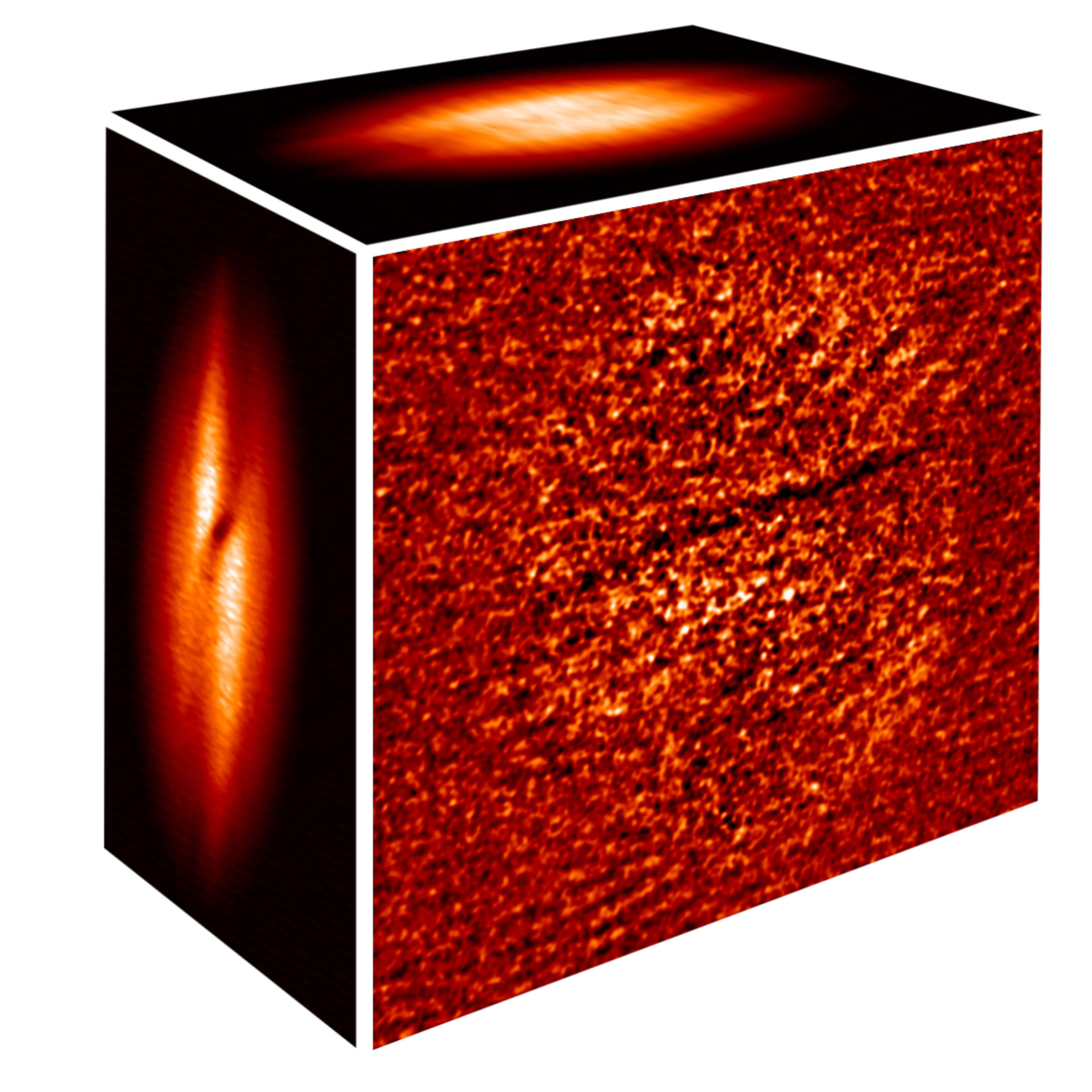}
\includegraphics[width = .24\textwidth]{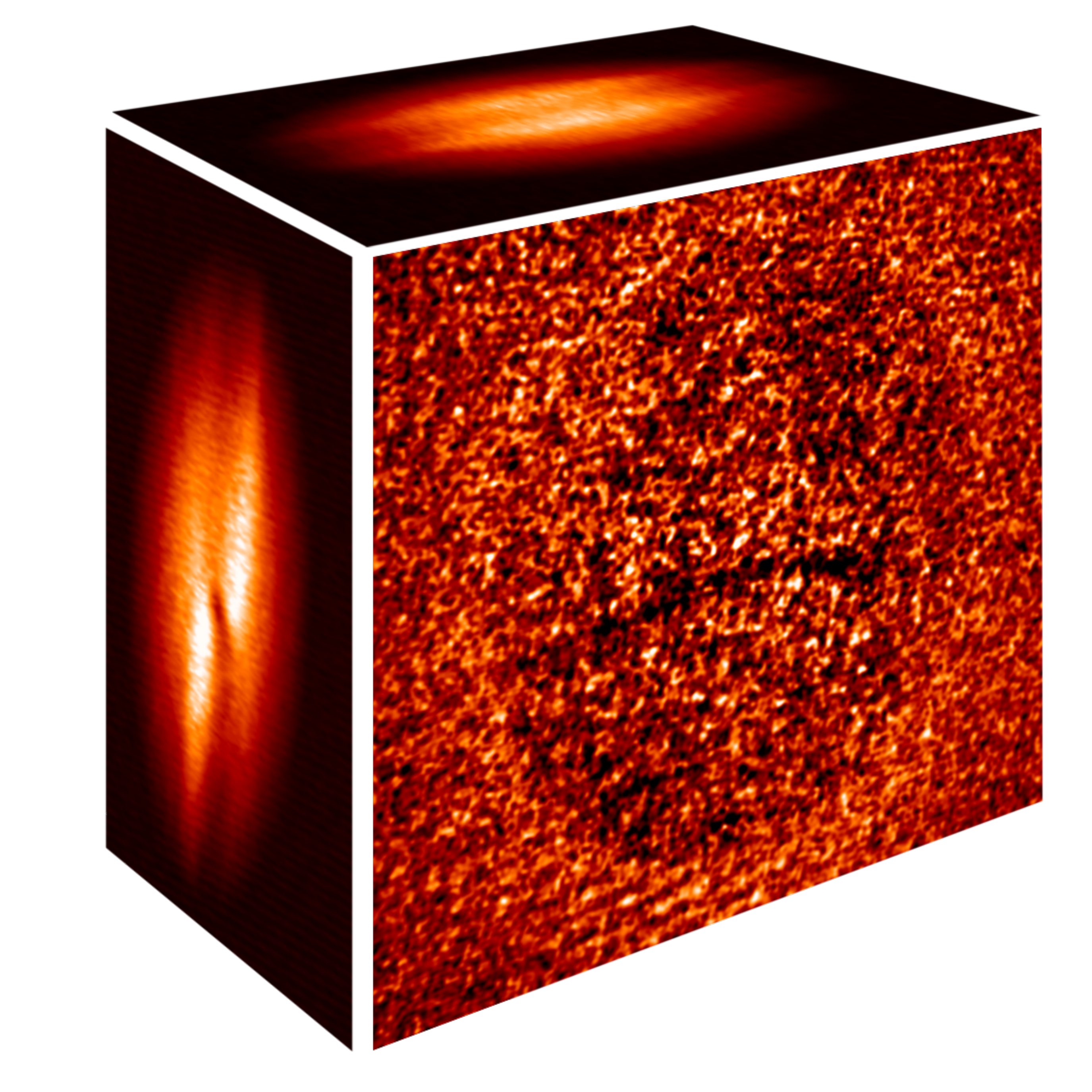}
\includegraphics[width = .24\textwidth]{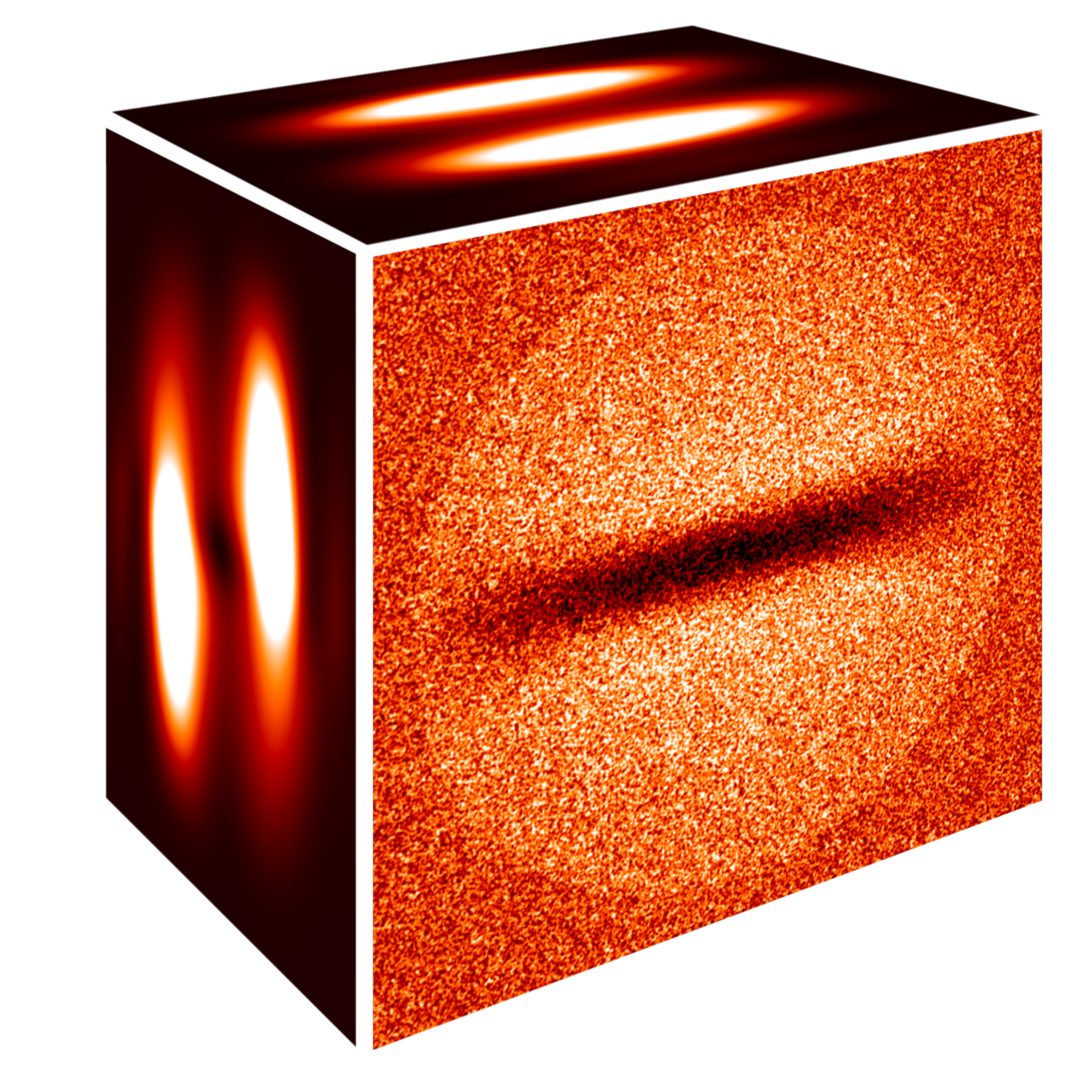}
\caption{Comparison between experimental images and numerical simulations. \textbf{(a-c)} Triaxial absorption imaging, after a free expansion time of 120 ms. The twist of the depletion in the density indicates a presence of a SV. The twist orientation indicates the vortex sign. \textbf{(a)} Vertical SV; \textbf{(b-c)} horizontal SV with opposite circulation; \textbf{(d)} GPE simulation with $\mu_{\rm theor} = 10\, \hbar \omega_\perp$ in case of a horizontal SV as in (c).}
\label{fig3:exp_density}
\end{figure}

Experimentally we investigate the density profile and the phase properties of the BEC after a time-of-flight of 120 ms ($\simeq 1.5$ axial oscillation periods). At such time the BEC has a pancake distribution. A vertical magnetic field gradient is applied after the trapping potential removal to levitate the BEC.

\textbf{Measurement of the density profile: triaxial absorption imaging}\\
We simultaneously image the density profile of our BEC from three orthogonal directions, the axial and two radial ones. The transverse size of the defects is of the order of the local healing length at any given time. After a long expansion time the defects are clearly visible from the radial direction. In the axial direction, on the other hand, the visibility is lower due to the integration of density along the whole BEC. In order to extract anyway useful information we then subtract the Thomas-Fermi profile fit from the data and plot the residuals (see Fig.  \ref{fig3:exp_density}).

From the axial direction we clearly see lines crossing the BEC, straight or curved, but always ending almost perpendicularly to the BEC outer surface. From the radial directions we either see straight lines or lines with a twist and a hollow core at the center. Combining these images we deduce the nature of the observed defect, a SV.

\textbf{Measurement of the phase profile: Bragg interferometry}\\
The triaxial density detection is quite rich of information and clearly leads us to identification our defects as SV. Nevertheless we implemented a homodyne Bragg interferometer in order to prove the 2$\pi$ phase winding around the expected vortex core.
\begin{figure}[!b]
\includegraphics[width = \textwidth]{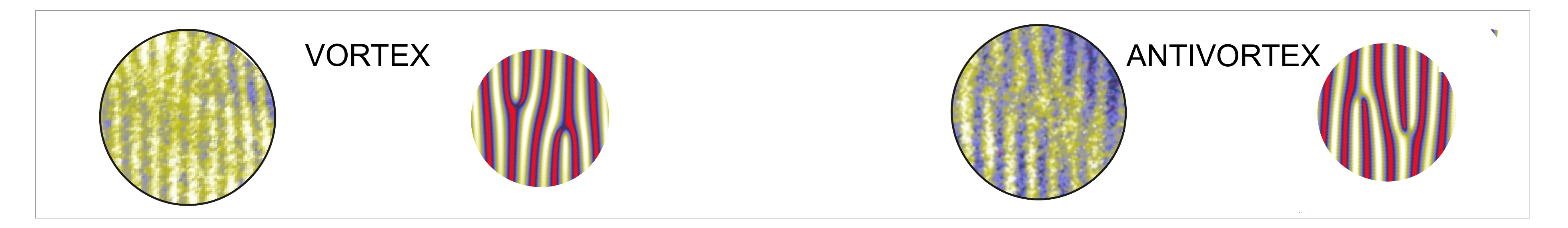}
\caption{Experimental and simulated interference pattern resulting after a Bragg interferometer applied on a BEC containing a vortex (left) or an antivortex (right). }
\label{fig4:exp_bragg}
\end{figure}
During time of flight we apply a sequence of two $\pi/2$ pulses. The first one splits the original BEC into two equal copies, one at rest and one moving with the Bragg momentum. The second pulse does the same on both copies and two pairs of shifted BEC copies are created. The pair at rest, as well as the pair that are moving, interfere with each other in the overlapping region and fringes can be detected. If the BEC phase was initially uniform then the fringe pattern results sinusoidal. In case of a vortex, instead, we see a pair of dislocation-antidislocation in the resulting phase pattern, as expected from simulation. Fig.  \ref{fig4:exp_bragg} shows the comparison between our experimental data and simulations. This proves that our defects do have circulation. Furthermore this confirms that the twist orientation is directly linked to the vortex sign \cite{Chevy01}.

\section{Conclusions}

In Ref. \cite{LamporesiKZM13}, the combination of  two radial images of an expanding condensate led to the identification of defects as solitons. A subsequent additional axial imaging and the interferometric phase measurement clearly demonstrate that the defects we observe are SVs. Our simulations show the crossover from a 1D regime, supporting solitons, to a 3D regime, supporting vortices. In the intermediate regime the planar density depletion and large phase gradient are clearly visible as signatures of SVs. Although in the experiment we are at the edge of the crossover region, with a small planar density depletion, we are still able to obtain striking evidences of the SV presence thanks to the amplification of the depletion contrast in expansion.

Since the defects are observed long after they are formed via the Kibble-Zurek mechanism occurring at the BEC transition and since we have no experimental access to the early dynamics of formation, it remains unknown whether the SV we see are decay products from initially formed solitons or if they form directly via KZM. With the parameters of our elongated condensates, the SV is expected soon to become the defect with the lowest energy \cite{Brand02} as the chemical potential increases during the cooling process. Further investigations are planned to clarify this issue, for instance by developing an appropriate imaging technique to observe the evolution of phase imprinted defects.\\

{\it Acknowledgements --} We thank J. Brand, C. Tozzo and M. Zwierlein for stimulating discussions. We acknowledge financial support by Provincia Autonoma di Trento and by PL-Grid Infrastructure.

\end{document}